\documentclass[aps, pre, reprint, amsmath, amssymb, superscriptaddress]{revtex4-2}

\usepackage{graphicx}
\graphicspath{{./figures}}
\usepackage{dcolumn}
\usepackage{bm}
\usepackage{lipsum}
\usepackage[colorlinks=true,linkcolor=blue,urlcolor=blue,citecolor=blue]{hyperref}
\usepackage[italicdiff]{physics}
\usepackage{bbold}
\usepackage{dsfont}

\begin{document}

\title{Information geometry of transitions between quantum nonequilibrium steady states}

\author{Artur M. Lacerda}
\email{machadoa@tcd.ie}
\affiliation{School of Physics, Trinity College Dublin, College Green, Dublin 2, D02K8N4, Ireland}

\author{Laetitia P. Bettmann}
\email{bettmanl@tcd.ie}
\affiliation{School of Physics, Trinity College Dublin, College Green, Dublin 2, D02K8N4, Ireland}

\author{John Goold}
\email{gooldj@tcd.ie}
\affiliation{School of Physics, Trinity College Dublin, College Green, Dublin 2, D02K8N4, Ireland}
\affiliation{Trinity Quantum Alliance, Unit 16, Trinity Technology and Enterprise Centre, Pearse Street, Dublin 2, D02YN67, Ireland}

\date{\today}
\begin{abstract}
In a transition between nonequilibrium steady states, the entropic cost associated with the maintenance of steady-state currents can be distinguished from that arising from the transition itself through the concepts of excess/housekeeping entropy flux and adiabatic/nonadiabatic entropy production. The thermodynamics of this transition is embodied by the Hatano-Sasa relation. In this letter, we show that for a slow transition between quantum nonequilibrium steady states the nonadiabatic entropy production is, to leading order, given by the path action with respect to a Riemannian metric in the parameter space which can be connected to the Kubo-Mori-Bogoliubov quantum Fisher information. We then demonstrate how to obtain minimally dissipative paths by solving the associated geodesic equation and illustrate the procedure with a simple example of a three-level maser. Furthermore, by identifying the quantum Fisher information with respect to time as a metric in state space, we derive an upper bound on the excess entropy flux that holds for arbitrarily fast processes.
\end{abstract}

\maketitle

\textbf{Introduction} - Nonequilibrium steady states (NESS) are fundamental to the study of systems maintained far from equilibrium, where persistent currents of energy, particles, or entropy flow continuously due to external driving forces or gradients~\cite{evans2008statistical}. Unlike equilibrium states, NESS embody dynamic stability, balancing inflows and outflows while breaking detailed balance, making them central to understanding dissipation, irreversibility, and entropy production~\cite{gaspard2022statistical}. These states are of paramount importance in a range of physical settings, from molecular motors~\cite{Seifert_2012} and nanoscale heat engines to quantum thermoelectric devices~\cite{benenti2017fundamental} and biological systems~\cite{Brown2020}, where steady-state currents drive functionality.

For quantum systems NESS are also key to understand dissipation and coherence-driven processes in quantum technologies~\cite{pekola2015towards,mitchison2019quantum}. In condensed matter, they are central to define the phenomenology of quantum transport~\cite{pekola2021quantum,landi2022nonequilibrium} and the emergence of diffusion~\cite{bertini2021finite} and nonequilibrium quantum critical phenomena~\cite{minganti2018spectral,fazio2024many}. Despite this significance, the study of transitions  between NESS, a critical process for driving systems under time-dependent protocols, remains under-explored in quantum thermodynamics. While in classical stochastic thermodynamics, transitions between NESS have been extensively studied as encapsulated by the Hatano-Sasa relation~\cite{Ooono1998,Hatano2001,Speck_2005, Ge2009, Ge2010, Esposito2010,Mandal2016}, providing deep insights into irreversibility and dissipation, only a few works have addressed the issue of NESS-to-NESS transitions in the quantum domain~\cite{Horowitz2014, Manzano2015, Manzano2018, Manzano2022}. By connecting both the excess entropy flux as well as the nonadiabatic entropy production to the geometry of the control parameter space, we provide new tools to quantify and optimize dissipation in quantum transitions.
\begin{figure}
    \centering
    \includegraphics{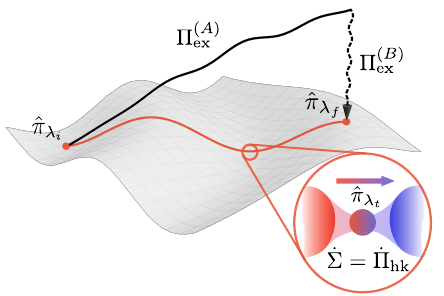}
    \caption{Schematic of a protocol driving the system between two nonequilibrium steady states. The system is initialized at a NESS $\hat{\pi}_{\lambda_i}$ specified by $\lambda_i$. In a quasistatic protocol connecting parameters $\lambda_i$ and $\lambda_f$ (orange line), the system remains in the manifold of NESS (grey surface), with $\hat{\pi}_{\lambda_t}$, at all times.  A  NESS is characterized by the entropy production rate $\dot{\Sigma}$ being balanced by the housekeeping flux rate $\dot{\Pi}_\mathrm{hk} $. For the finite-time transition (black line), we divide the ``excess'' entropy flux into a path dependent $\Pi^{A}_\mathrm{ex}$ and pure relaxation part $\Pi^{B}_\mathrm{ex}$.
    }
    \label{fig:manifold_diagram}
\end{figure}

The geometric perspective on thermodynamics is by now well established, as ~\cite{weinhold1975metric,ruppeiner1979thermodynamics,salamon1983thermodynamic} have provided profound insights into the costs of finite-time processes, with thermodynamic length and metrics offering tools to analyze dissipation and design efficient protocols~\cite{Crooks2007,Sivak2012}. This approach has been recently adopted to quantum thermodynamics~\cite{Scandi2019}, where it was used for process optimization ~\cite{abiuso2020geometric,rolandi2023finite} and to derive thermodynamic bounds~\cite{guarnieri2019thermodynamics, VanVuPRX2023}. These developments underscore the power of geometric frameworks to unify and generalize thermodynamic principles. Despite this progress, the geometry of transitions between NESS remains largely confined to the classical domain~\cite{Mandal2016}. In this Letter, we extend the geometric framework to analyze NESS-to-NESS transitions in the quantum regime. This approach both builds on and extends classical results~\cite{Mandal2016}, providing a unified quantum framework for understanding and optimizing dissipation in NESS transitions.  We show that the leading term of the excess entropy flux is given by the integral of a Riemannian metric along a path plus a boundary term. This boundary term is, as we discuss, precisely the excess flux of a pure relaxation process to the final NESS and the change in von Neumann entropy between initial and final NESS. In addition we provide a completely general upper bound on the excess entropy flux in NESS transitions which is based on recent works that explore the Fisher information with respect to time as applied to classical and quantum dynamics~\cite{PiresPRX2016, NicholsonPRE2018, ItoPRL2018, Nicholson2020, GarciaPRX2022, bringewatt2024, BettmannPRE2024}. 

\textbf{Thermodynamics} - We adopt a standard Lindblad master equation formalism where a time-dependent Liouvillian $\mathcal{L}_\lambda$ is specified by control parameters $\lambda_t$ and the system evolves according to  
\begin{equation}
    \label{eq:master_eq}
    \dot{\hat{\rho}}_t = \mathcal{L}_\lambda[\hat{\rho}_t].
\end{equation} 
We assume that, for a fixed $\lambda$, the system relaxes to a unique NESS $\hat{\pi}_\lambda$, satisfying $\mathcal{L}_\lambda[\hat{\pi}_\lambda]=0$. As is often done in phenomenological nonequilibrium thermodynamics~\cite{deGrootMazur1984} we split the entropy production rate $\dot{\Sigma}$ as
\begin{equation}
    \label{eq:second_law}
    \dv{S}{t} = \dot{\Sigma} - \dot{\Pi},
\end{equation}
where $S$ is the von Neumann entropy of the system and $\dot{\Pi}$ is the entropy flux  to the environment. For example, when the environment consists of multiple thermal baths at inverse temperature $\beta_\alpha$, $\dot{\Pi}=\sum_\alpha \beta_\alpha \langle \hat{J}^Q_\alpha\rangle$, where $\langle \hat{J}^Q_\alpha\rangle$ is the average heat current from the system to bath $\alpha.$

We adopt the approach of splitting $\dot{\Sigma}$ into adiabatic ($\dot{\Sigma}_\text{ad}$) and nonadiabatic ($\dot{\Sigma}_\text{na}$) contributions:
\begin{equation}
    \label{eq:non_ad_entropy}
    \dot{\Sigma} = \dot{\Sigma}_\text{ad} + \dot{\Sigma}_\text{na}, \quad \dot{\Sigma}_\text{na} = -\Tr[\dot{\hat{\rho}}_t (\log \hat{\rho}_t - \log \hat{\pi}_{\lambda_t})].
\end{equation}
This definition of $\dot{\Sigma}_\text{na}$ corresponds to the instantaneous version of Spohn’s entropy production relative to the instantaneous steady-state $\hat{\pi}_t$. It is nonnegative~\cite{Horowitz2014, Spohn1978} and vanishes only when the system reaches the steady state. When the steady state is thermal, $\dot{\Sigma}_\text{na}$ reduces to the entropy production described by the standard splitting~\eqref{eq:second_law}. For more general environments, where such a direct correspondence no longer holds, the difference between $\dot{\Sigma}$ and $\dot{\Sigma}_\text{na}$ naturally leads to our definition of $\dot{\Sigma}_\text{ad}$. In the steady state, $\dot{\Sigma}_\text{na}$ becomes zero, leaving $\dot{\Sigma}_\text{ad}$ as the only surviving term. This provides an interpretation for $\dot{\Sigma}_\text{ad}$ as the component of entropy production associated with the maintenance of the NESS rather than with relaxation toward it. As we discuss further in this paper, the slow driving expansion will provide further justification for this interpretation (see Supplemental Material~\cite{SM}).


For classical Markovian systems, both components can be defined at the level of trajectories, are nonnegative and individually satisfy fluctuation theorems~\cite{Esposito2010}. In quantum systems governed by a Lindblad equation, this is only true if further conditions are imposed on the jump operators, which imply that the steady state commutes with the system Hamiltonian \cite{Manzano2018,PhysRevE.103.052138}. Nevertheless, even when these conditions are not met, Eq.~\eqref{eq:non_ad_entropy} can still be used at the average level~\cite{Manzano2018}, and provides a robust quantifier for the additional dissipation in the process. While historically the first splitting to be proposed was that of the heat current ~\cite{Ooono1998, Hatano2001,Ge2009,Ge2010}, here we consider the splitting of the entropy flux into \textit{excess} and \textit{housekeeping flux}~\cite{Seifert_2012}.

The housekeeping entropy flux is equivalent, by construction, to the adiabatic component of the entropy production $\dot{\Pi}_\text{hk} = \dot{\Sigma}_\text{ad}$ as it is the only nonzero component in the steady state. This term quantifies the entropy supplied by the environment to maintain the NESS. The excess entropy flux is defined to be the difference $\dot{\Pi}_\text{ex} = \dot{\Pi} - \dot{\Pi}_\text{hk}$, which can be rewritten as (see Supplemental Material~\cite{SM})
\begin{equation}
    \label{eq:excess_ent_flux}
    \dot{\Pi}_\text{ex} = \Tr[\dot{\hat{\rho}}_t \log \hat{\pi}_{\lambda_t}].
\end{equation}
Notice that this quantity is zero in the steady state. With these definitions in hand, one can write a generalized Clausius inequality as~\cite{SM}
\begin{equation}
    \label{eq:modified_second_law}
    \dv{S}{t} + \dot{\Pi}_\text{ex}= \dot{\Sigma}_\text{na} \geq 0,
\end{equation}
which is a version of the celebrated Hatano-Sasa theorem \cite{Hatano2001}. Within this framework, a transition between NESS parallels a transition between equilibrium states, but with $\dot{\Sigma}_\text{na}$ playing the role of entropy production and $\dot{\Pi}_\text{ex}$ of the entropy flux.

\begin{figure*}
    \centering
    \includegraphics[width=\textwidth]{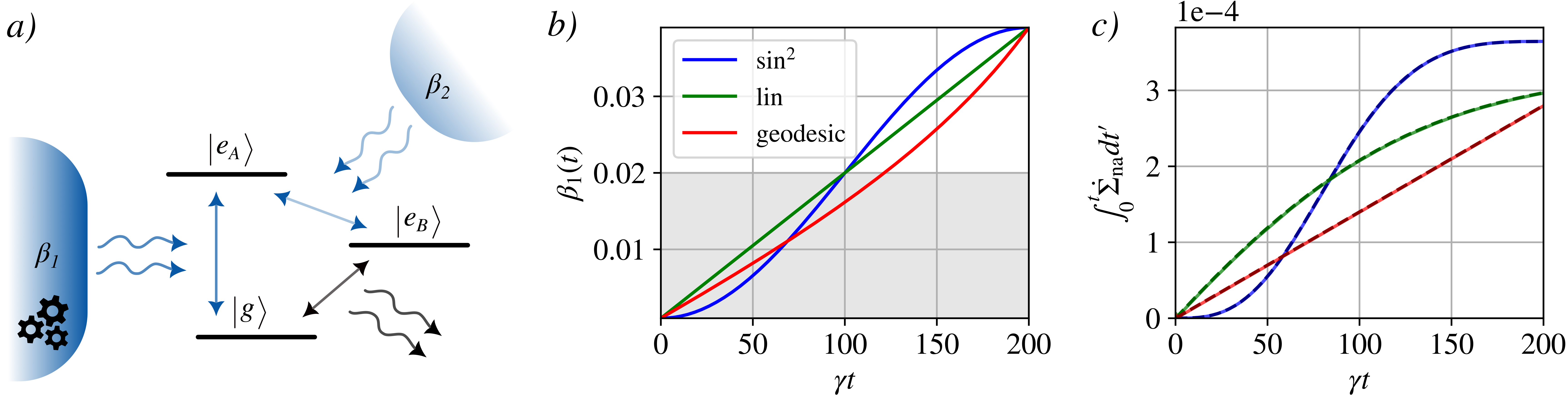}
    \caption{a) Schematic of the three-level maser operating as a quantum heat engine. The two reservoirs are modeled as thermal light sources with inverse temperatures $\beta_1$ and $\beta_2$, respectively. The absorption and emission of photons with frequencies $\omega_1 = \epsilon_A - \epsilon_g$ and $\omega_2 = \epsilon_A - \epsilon_B$ induces transitions between the respective levels. The population inversion, which enables stimulated emission at frequency  $\omega_3 = \epsilon_B - \epsilon_g$, which in this set-up is the work output, occurs between the levels $\ket{g}$ and $\ket{e_B}$. When $\omega_2/\omega_1 > \beta_1/\beta_2$ the machine operates as a heat engine (grey shaded in b)), whereas, in the complementary region, it operates as a refrigerator~\cite{Cangemi2024}. b) Geodesic protocol for $\beta_1$ as well as two `naive' protocols, defined in the Supplemental Material~\cite{SM}. c) In the slow-driving regime, $\langle\hat{\Phi}_{\lambda}\rangle\Big\vert_0^T + \int_0^T \,\dot\Pi_\text{ex} dt' \approx \int_0^T \,\dot{\Sigma}_\mathrm{na}dt' = \int_0^T \,\dot{\lambda}^T\zeta \dot{\lambda} dt' $. 
    Therefore, the total nonadiabatic entropy production is minimized for the optimal protocol with duration $T$ described by a geodesic connecting $\beta_1^0$ and $\beta_1^T$ in parameter space. The dashed lines represent the slow-driving approximation, while the solid lines are obtained from the exact expression.
 Parameters: $\gamma_1 = \gamma_2 = \gamma_3 = \gamma$, $\omega_1 = 50\gamma$, $\omega_2 = 10\gamma$, and $\gamma\beta_2= 0.1 $, $\gamma\beta^0_1 = 0.001$ to $\gamma\beta^T_1 = 0.039$ and $\gamma T = 200$.}
    \label{fig:3LM_all}
\end{figure*}

\textbf{Slow-driving} - Now let us consider a protocol in which the system is initially prepared in a NESS $\hat{\pi}_{\lambda_i}$ and the control parameters are slowly varied from $\lambda_i$ to $\lambda_f$ over a total time $T$, followed by a pure relaxation process to a new NESS $\hat{\pi}_{\lambda_f}$. The excess flux of the entire protocol $\Pi_\text{ex}$ is the sum of the excess flux accumulated over the parameter path and the pure relaxation process, which we denote by $\Pi_\text{ex}^{(A)}$
and $\Pi_\text{ex}^{(B)}$, respectively (see Fig.~\ref{fig:manifold_diagram}). Eq.~\eqref{eq:excess_ent_flux} motivates the introduction of the nonequilibrium potential \cite{Hatano2001}, which is an Hermitian operator defined by
\begin{equation}
    \hat{\Phi}_{\lambda} = -\log \hat{\pi}_{\lambda},
\end{equation}
such that $\dot{\Pi}_\text{ex} = -\Tr[\dot{\hat{\rho}}_t\hat{\Phi}_{\lambda}]$. Notice that $\dot{\Pi}_\text{ex}$ is not a complete differential, since $\hat{\pi}_\lambda$ also changes over time. By taking the derivative of $\ev{\hat{\Phi}_\lambda} = -\Tr[\hat{\rho}_t\log\hat{\pi}_{\lambda}]$, we can write
\begin{equation}
    \label{eq:excess_vs_non_eq_pot}
    \dot{\Pi}_\text{ex} = -\dv{\langle\hat{\Phi}_\lambda\rangle}{t} - \Tr[\hat{\rho}_t \dv{\log\hat{\pi}_\lambda}{t}].
\end{equation}
For a pure relaxation protocol, where $\hat{\pi}_\lambda$ is fixed, the second term vanishes. Therefore,
\begin{equation}
    \label{eq:excess_pure_relax}
    \Pi_\text{ex}^{(B)} = \Tr[(\hat{\rho}_T - \hat{\pi}_{\lambda_f})\hat{\Phi}_{\lambda_f}],
\end{equation}
which is the difference in expectation value of the nonequilibrium potential computed at $\hat{\rho}_T$ and $\hat{\pi}_{\lambda_f}$.

Let us now focus on $\Pi_\text{ex}^{(A)}$. Following \cite{Mandal2016, CavinaPRL2017}, we introduce a small parameter $\epsilon$ proportional to the driving speed, such that $\epsilon \propto 1/T$. For sufficiently slow driving, the system will remain close $\hat{\pi}_{\lambda_t}$, up to a small correction. Consider the expansion
\begin{equation}
    \label{eq:correction_ness}
    \hat{\rho}_t = \hat{\pi}_{\lambda} + \delta\hat{\rho}_t,
\end{equation}
where $\Tr[\delta\hat{\rho}_t]=0$, by the preservation of trace. Inserting this expansion into Eq.~\eqref{eq:master_eq} and using that $\mathcal{L}_\lambda[\hat{\pi}_\lambda]=0$, we obtain a differential equation for $\delta\hat{\rho}_t$~\cite{Scandi2019, Mandal2016, CavinaPRL2017}:
\begin{equation}
    \label{eq:time_evo_correction_ness}
    \left(\mathcal{L}_\lambda-\dv{t}\right)[\delta\hat{\rho}_t] = \dv{\hat{\pi}_\lambda}{t}.
\end{equation}
Eq.~\eqref{eq:time_evo_correction_ness} can be formally solved by the introduction of the Drazin inverse operator~\cite{Gower1972}, which is defined as the unique operator satisfying the properties (i) $\mathcal{L}_\lambda\mathcal{L}_\lambda^+[\hat{A}] = \mathcal{L}_\lambda^+\mathcal{L}_\lambda[\hat{A}] = \hat{A} - \hat{\pi}_\lambda \Tr[\hat{A}]$, (ii) $\mathcal{L}_\lambda^+[\hat{\pi}_\lambda]=0$ and (iii) $\Tr[\mathcal{L}_\lambda^+[\hat{A}]]=0$. It can be shown that the Drazin inverse has an integral representation given by~\cite{Gower1972}
\begin{equation}
    \label{eq:drazin_integral}
    \mathcal{L}_\lambda^+[\hat{A}] := \int_0^\infty d\tau\, e^{\mathcal{L}_\lambda\tau}\qty(\hat{\pi}_\lambda \Tr[\hat{A}]-\hat{A}).
\end{equation}

By applying $\mathcal{L}_\lambda^+$ on both sides, we obtain the expansion (see Supplemental Material~\cite{SM} for details)
\begin{equation}
    \label{eq:expansion_rho}
    \hat{\rho}_t =\hat{\pi}_\lambda + \sum_{n=1}^\infty\left(\mathcal{L}_\lambda^+\dv{t}\right)^n\hat{\pi}_\lambda.
\end{equation}
Notice that the $n^\text{th}$ term in this series contains an $n^\text{th}$-order derivative of $\hat{\pi}_\lambda$, and is thus proportional to $\epsilon^n$, which means that Eq.~\eqref{eq:expansion_rho} can be seen as an expansion in powers of $\epsilon$.

As an important consistency check, Eq.~\eqref{eq:expansion_rho} can be used to obtain an expression for $\dot{\Pi}_\text{ex}$ in the quasistatic limit. In this case, the final state is already a NESS, thus $\Pi_\text{ex}^{(B)}=0$ and $\dot{\Pi}_\text{ex}=\dot{\Pi}_\text{ex}^{(A)}$. By taking the derivative of Eq.~\eqref{eq:expansion_rho} and keeping only terms up to first order in $\epsilon$, we see that $\dot{\hat{\rho}}_t = \dot{\hat{\pi}}_\lambda + \mathcal{O}(\epsilon^2)$. Inserting this result into Eq.~\eqref{eq:excess_ent_flux} we obtain
\begin{equation}
    \dot{\Pi}_\text{ex} = -\dv{S}{t} + \mathcal{O}(\epsilon^2).
\end{equation}
Here, since $\hat{\rho}_t$ coincides with $\hat{\pi}_{\lambda_t}$,  $S(\hat{\pi}_\lambda) = S(\hat{\rho}_t).$ By integrating over the path, we get the generalized Clausius equality $\Delta S + \int_0^T dt\,\dot{\Pi}_\text{ex} \overset{\text{qs}}{=} 0.$

To go beyond the quasistatic limit, we insert the expansion \eqref{eq:expansion_rho} into Eq.~\eqref{eq:excess_vs_non_eq_pot}, which results in
\begin{equation}
    \label{eq:excess_flux_A_expansion}
    \dot{\Pi}_\text{ex} = -\dv{\langle\hat{\Phi}_{\lambda}\rangle}{t} - \sum_{n=1}^\infty \Tr[\dv{\log\hat{\pi}_\lambda}{t}\qty(\mathcal{L}_\lambda^+\dv{t})^n\hat{\pi}_\lambda],
\end{equation}
where we used the fact that $\Tr[\hat{\pi}_\lambda \dv{\log\hat{\pi}_\lambda}{t}]=0$~\cite{Horowitz2014} to start the sum for $n=1$. By integrating over the protocol and keeping only the leading order term, we obtain
\begin{equation}
\label{eq: excess_SD}
    \Pi_\text{ex}^{(A)} =  -\langle\hat{\Phi}_{\lambda}\rangle\Big\vert_0^T -\int_0^T dt\,\Tr[\dv{\log\hat{\pi}_\lambda}{t}\mathcal{L^+_\lambda}\dv{\hat{\pi}_\lambda}{t}] + \mathcal{O}(\epsilon^2),
\end{equation}
where $\langle\hat{\Phi}_{\lambda}\rangle\big\vert_0^T = \Tr[\hat{\rho}_T\hat{\Phi}_{\lambda_f}]-\Tr[\hat{\pi}_{\lambda_i}\hat{\Phi}_{\lambda_i}]$. In the Supplemental Material~\cite{SM}, we show that the boundary term can be expressed as
\begin{equation}
    \label{eq:boundary_term}
    \langle\hat{\Phi}_{\lambda}\rangle\Big\vert_0^T = \Delta S(\hat{\pi}_{\lambda}) + \Pi_\text{ex}^{(B)},
\end{equation}
where $\Delta S(\hat{\pi}_{\lambda}) = S(\hat{\pi}_{\lambda_f})-S(\hat{\pi}_{\lambda_i})$. Notice that this quantity depends only on the endpoints of the protocol. Therefore, by incorporating the full relaxation process, we can proceed without assuming that the velocity of the protocol is zero at the endpoints, extending the results of~\cite{Mandal2016}. Using the chain rule, we can write
\begin{equation}
    \label{eq:first_correction_chain_rule}
    \Pi_\text{ex} = -\Delta S(\hat{\pi}_\lambda)+\sum_{\mu\nu}\int_0^T dt\,\dot{\lambda}_\mu\zeta_{\mu\nu}\dot{\lambda}_\nu,
\end{equation}
where $\zeta_{\mu\nu} = \frac{1}{2}(\xi_{\mu\nu} + \xi_{\nu\mu})$ is the symmetric part of
\begin{equation}
    \label{eq:friction_tensor_unsymmetric}
    \xi_{\mu\nu} =-\Tr[\pdv{\log\hat{\pi}_\lambda}{\lambda_\nu}\mathcal{L^+_\lambda}\pdv{\hat{\pi}_\lambda}{\lambda_\mu}].
\end{equation}
Furthermore, in the Supplemental Material~\cite{SM} we derive the Green-Kubo relation for $\xi$, showing that
\begin{equation}
    \label{eq:green_kubo_elements}
    \xi_{\mu\nu} = \int_0^\infty d\tau\,\langle \hat{F}^\mu, \hat{F}^\nu(\tau)\rangle_{\hat{\pi}_\lambda},
\end{equation}
where $\ev{\cdot,\cdot}_{\hat{\pi}_\lambda}$ stands for the Kubo-Mori-Bogoliubov (KMB) inner product~\cite{Petz1993}, defined as $\langle \hat{A}, \hat{B}\rangle_{\hat{\pi}_\lambda} = \int_0^1 ds\, \Tr[\hat{\pi}_\lambda^s \hat{A}^\dagger \hat{\pi}_\lambda^{1-s}\hat{B}]$ and $\hat{F}^\mu(\tau)$ is the result of time-evolving the logarithmic derivative operator, which is defined as $\hat{F}^\mu = \pdv{\log \hat{\pi}_\lambda}{\lambda_\mu}$, in the Heisenberg picture.

Following \cite{Sivak2012, Mandal2016}, the friction tensor $\zeta$ can be related to the KMB quantum Fisher information matrix (QFI), which is defined as
\begin{equation}
    \label{eq:QFI_matrix}
    \mathcal{I}_{\mu\nu} = \ev{\hat{F}^\mu, \hat{F}^\nu}_{\hat{\pi}_\lambda},
\end{equation}
by the introduction of the generalized integral relaxation time
\begin{equation}
    \tau_{\mu\nu} = \int_0^\infty d\tau\, \frac{\langle \hat{F}^\mu, \hat{F}^\nu(\tau)\rangle_{\hat{\pi}_\lambda}}{\langle \hat{F}^\mu, \hat{F}^\nu\rangle_{\hat{\pi}_\lambda}},
\end{equation}
which is the characteristic time for $\hat{F}^\mu$ and $\hat{F}^\nu$ to become uncorrelated~\cite{Garanin1990}. By inspection, we see that $\xi_{\mu\nu} = \tau_{\mu\nu}\mathcal{I}_{\mu\nu}$, therefore
\begin{equation}
\label{eq:zeta}
    \zeta_{\mu\nu} = \frac{1}{2}\qty(\tau_{\mu\nu}  \mathcal{I}_{\mu\nu} + \tau_{\nu\mu}\mathcal{I}_{\nu\mu}).
\end{equation}

In the Supplemental Material~\cite{SM}, we show that $\zeta(\lambda)$ is a positive semi-definite matrix. Since it is also symmetric by construction, it establishes a Riemannian metric on the space of NESS parameterized by $\lambda$.

We note that our derivation generalizes the classical result of~\cite{Mandal2016}. In fact, up until Eq.~\eqref{eq:friction_tensor_unsymmetric}, our results are in complete analogy with those of~\cite{Mandal2016}. However, the computation of the Green-Kubo elements [\eqref{eq:green_kubo_elements}] is significantly different, due to the noncommuting nature of the operators. In our derivation, the order of the operators forces the choice of the KMB inner product, while classically there exists only one choice. Therefore, our considerations of the boundary term also hold classically, by replacing the KMB inner product with the classical correlator, which allows for obtaining the metric in protocols with non-vanishing velocities at the endpoints.

In the Supplemental Material~\cite{SM}, we also show that the action of this metric is connected to nonadiabatic entropy production by
\begin{equation}
\label{eq:action_metric_non_ad_ent}
\int_0^T\mathrm{d}t\,\sum_{\mu\nu}\dot{\lambda}_\mu\zeta_{\mu\nu}\dot{\lambda}_\nu = \int_0^T\mathrm{d}t\,\dot{\Sigma}_{\text{na}} + \mathcal{O}(\epsilon^2),
\end{equation}
 By the Cauchy-Schwarz inequality, this implies that the geometric action is bounded from below by the square of the path length, $l =  \int_0^T\mathrm{d}t\,\sqrt{\sum_{\mu\nu}\dot{\lambda}_\mu\zeta_{\mu\nu}\dot{\lambda}_\nu}$, divided by the protocol duration $T$, so that,  ${\Sigma}_{\text{na}}\geq l^2/T$. Importantly, equality holds only for a geodesic. This motivates seeking such minimal-length paths, which in the slow-driving regime, correspond to minimally dissipative protocols.

\textbf{Geodesics and optimal protocols} - 
The geodesic path is the solution to the geodesic equation
\begin{equation}
\label{eq: geodesic_eq}
\frac{d^2\lambda^\alpha}{dt^2} + \Gamma_{\beta \gamma}^\alpha \frac{d\lambda^\beta }{dt} \frac{d\lambda^\gamma}{dt} =0
\end{equation}
where the Christoffel symbols 
\begin{equation}
\label{eq: christoffel}
    \Gamma^\alpha_{\beta \gamma} = \frac{1}{2} \zeta^{\alpha \delta} \left(\frac{d\zeta_{\gamma \delta}}{d\lambda^\beta } + \frac{d\zeta_{\beta \delta}}{d\lambda^\gamma} -\frac{d\zeta_{\beta \gamma}}{d\lambda^\delta} \right),
\end{equation}
depend on the friction tensor $\zeta$, which we previously identified as a metric.
An analytic expression for the optimal protocol is typically unavailable, requiring reliance on numerical methods to determine the geodesic. 

To illustrate our results, we analyze dissipation during the execution of a slow-driving protocol using a three-level maser (TLM) as the working medium of a quantum thermal machine, as shown in Fig.~\ref{fig:3LM_all}a). The protocols we consider connect two distinct operational regimes: the TLM initially operates as a heat engine and then transitions to a regime where it functions as a refrigerator~\cite{Cangemi2024}.
To this end, we control the inverse temperature $\beta_1(t)$ (see Fig.~\ref{fig:3LM_all}a) and b)). Leveraging the fact that the nonadiabatic entropy production can be expressed as the path action for the metric $\zeta$, we obtain a minimally dissipative protocol by finding a geodesic path in parameter space (see Eqs.~\eqref{eq: geodesic_eq} and~\eqref{eq: christoffel}). This procedure, and its simplification for a single-parameter protocol, is discussed in detail in the Supplemental Material~\cite{SM}. The resulting geodesic protocol for $\beta_1(t)$ is shown in Fig.~\ref{fig:3LM_all}b) alongside two naive protocols (see Supplemental Material~\cite{SM}). We observe excellent agreement between the slow-driving approximation and the exact integrated excess flux, as illustrated by the solid and dashed lines in Fig.~\ref{fig:3LM_all}c).
As expected, the nonadiabatic entropy production is minimized by the optimal protocol represented by the geodesic.

\textbf{Arbitrary-speed transitions} -
To move beyond the limitations of slow driving, we now explore dissipation dynamics independent of driving speed. While one approach involves incorporating higher-order terms in the expansion of the instantaneous state, it is unlikely to yield broadly applicable results. Instead, by transitioning from the geometric space of control parameters to a time-parameterized geometric space~\cite{PiresPRX2016, NicholsonPRE2018, ItoPRL2018, Nicholson2020, GarciaPRX2022, bringewatt2024, BettmannPRE2024} (see Supplemental Material~\cite{SM}), we link the excess entropy flux to the statistical length, providing new insights into fast-driving dynamics.
To this end, we may make use of the fact that any member of the family of QFI with respect to the time parameter, 
\begin{equation}
I^f_Q(t) = \sum_{x, y} \frac{|\partial_t \hat{\rho}_{xy}(t)|^2}{p_x(t) f(p_y(t)/p_x(t))}.
\end{equation}
with the corresponding standard monotone functions $f$ (see Supplemental Material~\cite{SM}), spectrally decomposed density matrix $\hat{\rho} = \sum_x p_x \ket{x}\bra{x}$ and $d\hat{\rho}_{xy} := \bra{x}d\hat{\rho}\ket{y}$, plays the role of metric in state space~\cite{petz_introduction_2011, PETZ1996}. This means that for small changes in $t$ we can assign an infinitesimal length element
 \begin{equation}
     ds_f^2 = \frac{1}{4} I_Q^f dt^2,
 \end{equation}
 which lets us identify $ ds_f/dt = 1/2 \sqrt{I_Q^f} =: 1/2 v_f$ as speeds in state space.
Building on this geometric connection, in~\cite{Pintos2022,bringewatt2024} a generalized speed limit on an arbitrary observable $\hat{A}$ in terms of $v_f$, resembling the celebrated Mandelstam-Tamm time-energy uncertainty relation~\cite{Mandelstam1991},  
 \begin{equation}
     \vert \dot{a}\vert \leq \sigma^f_{\hat{\rho}} [\hat{A}]v_f,
 \end{equation}
where $\dot{a}= \Tr \left[A\dot{\rho}\right]$, $\sigma^2_{\hat{\rho}}[\hat{A}] = \Tr \left[\hat{A}_0 L_{\hat{\rho} }f(L_{\hat{\rho}}^{-1} R_{\hat{\rho}})(\hat{A}_0)\right]$, with $\hat{A}_0=\hat{A}-\Tr\left[\hat{\rho} \hat{A}\right]$, is a generalized variance that reduces to the standard one for the commonly used QFI based on the symmetric logarithmic derivative (SLD), and super-operators $L_{\hat{\rho}}(\cdot ) = \hat{\rho}(\cdot)$ and $R_{\hat{\rho}}(\cdot ) = (\cdot)\hat{\rho}$.
The observable of interest in the context of excess entropy flux is the nonequilibrium potential, ${\hat{\Phi}} = -\log{\hat{\pi}_\lambda}$, since $\dot{\Pi}_\text{ex} = -\Tr\left[\dot{\hat{\rho}}{\hat{\Phi}}\right]$. This connection allows us to derive a geometric bound on the excess entropy flux
\begin{equation}
    \vert \dot{\Pi}_\text{ex} \vert  
    \leq \sigma^f_{\hat{\rho}} [\hat{\Phi}] v_f.
 \end{equation}
  We emphasize that this bound holds for arbitrary-speed transitions.
 Upon integration, one finds
 \begin{equation}
 \begin{split}
      \int_0^T dt \,\vert \dot{\Pi}_\text{ex} \vert  
    &\leq \int_0^T dt  \,\sigma^f_{\hat{\rho}} [\hat{\Phi}] \sqrt{I_Q^f} \\
 \end{split}
 \end{equation}
 and thus a bound on the excess entropy flux. 
 Alternatively, the geometric interpretation of $v_f$ as a speed in state space allows its integration to yield the statistical length 
 \begin{equation}
 \label{eq: arb_speed_bound3}
    \int_0^T dt  \,\frac{\vert \dot{\Pi}_\text{ex} \vert }{\sigma^f_{\hat{\rho}} [\hat{\Phi}]}
    \leq \int_0^T dt  \, v_f = 2 \ell_f,
 \end{equation}
 where $\ell_f = \int_\gamma ds_f$.
 These bounds reveal how the system's path $\gamma$ in state space during transitions between different NESS, or, in fact, for arbitrary Lindblad dynamics with a unique fixed point, imposes geometric constraints on dissipation dynamics. Notably, the upper bound in the integrated speed limit \eqref{eq: arb_speed_bound3} is most stringent for a geodesic in state space, whose length can be determined analytically from a closed-form expression when the QFI metric is given by either the SLD-based quantum Fisher information or the Wigner-Yanase skew information~\cite{Wigner1963}.
 
\textbf{Conclusion -}
In this Letter, we have extended the geometric framework of thermodynamics to slow transitions between quantum nonequilibrium steady states (NESS), revealing the critical role of the Kubo-Mori-Bogoliubov quantum Fisher information in quantifying and optimizing dissipation. By deriving geodesics in the parameter space, we identified paths of minimal nonadiabatic entropy production, with potential applications in quantum control. In addition, we were also able to derive a geometric upper bound on the excess flux by switching from the space of control parameters to time --- a result that holds for arbitary speed NESS transitions, and more generally, for arbitrary Lindblad dynamics with a unique fixed point. Ongoing work focuses on applying these insights to quantum dissipative phase transitions, aiming to uncover new universality classes in nonequilibrium quantum systems.

\textbf{Acknowledgments} - J.G. is supported by a SFI - Royal Society University Research Fellowship and funding from Research Ireland under the QUAMNESS grant which is an EPRSC joint funding initiative. L.P.B. and J.G. also acknowledge Research Ireland for support through the Frontiers for the Future project. The authors thank G. Guarnieri, A. Rolandi,  M.~T. Mitchison and F. Binder for useful discussions.

\bibliographystyle{apsrev4-2}
\bibliography{references}

\onecolumngrid

\appendix

\newpage
\begin{center}
{\textbf{Supplementary Material}}
\end{center}

\section{\label{appendix:splitting_excess_hk}Excess/housekeeping entropy flux}
In this appendix we derive Hatano-Sasa's relation [Eq.~\eqref{eq:modified_second_law}] and Eq.~\eqref{eq:excess_ent_flux} from the splittings of the entropy production and entropy flux. We start by inserting the splittings $\dot{\Sigma}=\dot{\Sigma}_\text{ad} + \dot{\Sigma}_\text{na}$ and $\dot{\Pi}= \dot{\Pi}_\text{ex} + \dot{\Pi}_\text{hk}$ in the expression for the second law \eqref{eq:second_law}, which results in
\begin{equation}
    \dv{S}{t} = \qty(\dot{\Sigma}_\text{ad} + \dot{\Sigma}_\text{na}) - \qty(\dot{\Pi}_\text{ex} + \dot{\Pi}_\text{hk}).
\end{equation}
By definition, $\dot{\Pi}_\text{hk} = \dot{\Sigma}_\text{ad}$, so these two terms cancel. Therefore, we arrive at
\begin{equation}
    \dv{S}{t} + \dot{\Pi}_\text{ex} = \dot{\Sigma}_\text{na} = -\Tr[\dot{\hat{\rho}}_t (\log \hat{\rho}_t - \log \hat{\pi}_{\lambda_t})]\geq 0.
\end{equation}
which is Eq.~\eqref{eq:modified_second_law} of the main text. Isolating $\dot{\Pi}_\text{ex}$, we obtain
\begin{align}
    \begin{split}
        \dot{\Pi}_\text{ex} &= -\dv{S}{t}-\Tr[\dot{\hat{\rho}}_t(\log\hat{\rho}_t - \log\hat{\pi}_{\lambda_t})]\\
        &= \Tr[\dot{\hat{\rho}}_t\log\hat{\rho}_t]-\Tr[\dot{\hat{\rho}}_t(\log\hat{\rho}_t - \log\hat{\pi}_{\lambda_t})]\\
        &=\Tr[\dot{\hat{\rho}}\log\hat{\pi}_{\lambda_t}],
    \end{split}
\end{align}
which is Eq.~\eqref{eq:excess_ent_flux} of the main text.

\section{\label{appendix:slow_driving_expansion}Slow-driving expansion}

In this section, we derive the slow-driving expansion for $\hat{\rho}_t$. Applying $\mathcal{L}_\lambda^+$ on both sides of Eq.~\eqref{eq:time_evo_correction_ness} results in
\begin{equation}
    \mathcal{L}_\lambda^+\mathcal{L}_\lambda[\delta \hat{\rho}_t]-\mathcal{L}_\lambda^+\dv{(\delta \hat{\rho}_t)}{t}=\mathcal{L}_\lambda^+ \dv{\hat{\pi}_\lambda}{t}.
\end{equation}
By property (i) of the Drazin inverse, $\mathcal{L}_\lambda^+\mathcal{L}_\lambda[\delta \hat{\rho}_t] = \delta \hat{\rho}_t - \hat{\pi}_\lambda \Tr[\delta \hat{\rho}_t]$, therefore
\begin{equation}
    \mathbb{1}-\hat{\pi}_\lambda\Tr[\delta \hat{\rho}_t]-\mathcal{L}_\lambda^+\dv{(\delta \hat{\rho}_t)}{t}=\mathcal{L}_\lambda^+ \dv{\hat{\pi}_\lambda}{t}.
\end{equation}
Using the fact that $\Tr[\delta \hat{\rho}_t]=0$, we get
\begin{equation}
     \bigg(\mathbb{1}-\mathcal{L}_\lambda^+\dv{t}\bigg)[\delta\hat{\rho}_t] = \mathcal{L}_\lambda^+\dv{\hat{\pi}_\lambda}{t}.
\end{equation}
This equation can be formally solved by inverting the operator $\left(\mathbb{1}-\mathcal{L}_\lambda^+\dv{t}\right)$:
\begin{equation}
     \delta\hat{\rho}_t = \bigg(\mathbb{1}-\mathcal{L}_\lambda^+\dv{t}\bigg)^{-1}\mathcal{L}_\lambda^+\dv{\hat{\pi}_\lambda}{t}.
\end{equation}
The inversion is done via the expansion
\begin{equation}
    \bigg(\mathbb{1}-\mathcal{L}_\lambda^+\dv{t}\bigg)^{-1}=\sum_{n=0}^\infty\bigg(\mathcal{L}_\lambda^+\dv{t}\bigg)^n,
\end{equation}
which follows by formally treating the expansion $(1-x)^{-1} = 1 + x + x^2 + ...$ as power series generating function. Therefore,
\begin{equation}
    \hat{\rho}_t =\sum_{n=0}^\infty\bigg(\mathcal{L}_\lambda^+\dv{t}\bigg)^n\hat{\pi}_\lambda.
\end{equation}

\section{\label{appendix:boundary_term}Boundary term}

In the following we derive Eq.~\eqref{eq:boundary_term} of the main text. Using that $\hat{\rho}_T = \hat{\pi}_{\lambda_T} + \delta\hat{\rho}_T$, Eq.~\eqref{eq:excess_pure_relax} can be rewritten as
\begin{equation}
    \Pi_\text{ex}^{(B)} = -\Tr[\delta\hat{\rho}_T\hat{\Phi}_{\lambda_f}],
\end{equation}
Since $\hat{\rho}_0=\hat{\pi}_{\lambda_i}$, we then have that
\begin{align}
    \begin{split}
        \langle&\hat{\Phi}_{\lambda}\rangle\Big\vert_0^T = \Tr[\hat{\rho}_T \log\hat{\pi}_{\lambda_f}]-\Tr[\hat{\pi}_{\lambda_i}\log\hat{\pi}_{\lambda_i}]\\
        &= \Tr[\hat{\pi}_{\lambda_f} \log\hat{\pi}_{\lambda_f}]-\Tr[\hat{\pi}_{\lambda_i}\log\hat{\pi}_{\lambda_i}] -\Tr[\delta\hat{\rho}_T \log\hat{\pi}_{\lambda_f}]\\
        &= -\Delta S(\hat{\rho}_{\lambda}) - \Pi_\text{ex}^{(B)},
    \end{split}
\end{align}
which completes the proof.

\section{\label{appendix:green_kubo}Green-Kubo relations}

In this section we derive a Green-Kubo relation for the friction tensor $\zeta$. Using the integral form of the Drazin inverse, see Eq.~\eqref{eq:drazin_integral}, in Eq.~\eqref{eq:friction_tensor_unsymmetric}, we get
\begin{align}
        \xi_{\mu\nu} &= -\int_0^\infty d\tau\Tr[\pdv{\log\hat{\pi}_\lambda}{\lambda_\nu}e^{\mathcal{L_\lambda}\tau}\qty(\hat{\pi}_{\lambda}\Tr[\pdv{\hat{\pi}_\lambda}{\lambda_\mu}]-\pdv{\hat{\pi}_\lambda}{\lambda_\mu})]\\
        &=-\int_0^\infty d\tau\Tr[\pdv{\log\hat{\pi}_\lambda}{\lambda_\nu}e^{\mathcal{L_\lambda}\tau}\qty(\pdv{\hat{\pi}_\lambda}{\lambda_\mu})]\\
        &= -\int_0^\infty d\tau\Tr[\hat{F}^\nu e^{\mathcal{L_\lambda}\tau}\qty(\pdv{\hat{\pi}_\lambda}{\lambda_\mu})]
\end{align}
where in the second equality we used that $\Tr[\pdv{\hat{\pi}_{\lambda}}{\lambda_\mu}]=0$, by conservation of the trace. Recall that in the Heisenberg picture, operators evolve according to $\hat{F}^\nu(\tau) = e^{\mathcal{L}^\dagger_\lambda \tau}[\hat{F}^\nu]$,
where $\mathcal{L}^\dagger$ is formally defined as the adjoint of $\mathcal{L}$ under the Hilbert-Schmidt inner product, which means that $\Tr[\hat{A}\,\mathcal{L}(\hat{B})]=\Tr[\mathcal{L}^\dagger(\hat{A})\hat{B}]$, for $\hat{A},\hat{B}$ Hermitian matrices. Therefore, $\Tr[\hat{F}^\nu e^{\mathcal{L_\lambda}\tau}\qty(\pdv{\hat{\pi}_\lambda}{\lambda_\mu})]=\Tr[e^{\mathcal{L^\dagger_\lambda}\tau}(\hat{F}^\nu) \pdv{\hat{\pi}_\lambda}{\lambda_\mu}]$, then
\begin{equation}
    \xi_{\mu\nu} =-\int_0^\infty d\tau\Tr[\hat{F}^\nu(\tau)\pdv{\hat{\pi}_\lambda}{\lambda_\mu}].
\end{equation}

To proceed, we notice that the logarithmic derivative operator $\hat{F}^\nu$ satisfies the property \cite{Hayashi2002}
\begin{equation}
    \label{eq:green_kubo_derivation}
    \pdv{\hat{\pi}_\lambda}{\lambda_\mu} = \int_0^1 ds\, 
    \hat{\pi}_\lambda^s \hat{F}^\mu\hat{\pi}_\lambda^{1-s},
\end{equation}
which can also be used as an implicit definition of $\hat{F}^\mu$. Using this expression in Eq.~\eqref{eq:green_kubo_derivation}, we get
\begin{align}
    \begin{split}
        \xi_{\mu\nu} &= -\int_0^\infty d\tau\int_0^1 ds\Tr[\hat{F}^\mu(\tau) \hat{\pi}_\lambda^s\hat{F}^\nu\hat{\pi}_\lambda^{1-s}]\\
        &= -\int_0^\infty d\tau\int_0^1 ds\Tr[ \hat{\pi}_\lambda^s\hat{F}^\nu\hat{\pi}_\lambda^{1-s}\hat{F}^\mu(\tau)]\\
        &= \int_0^\infty d\tau\,\langle \hat{F}^\mu(0), \hat{F}^\nu(\tau)\rangle_{\hat{\pi}_\lambda},
    \end{split}
\end{align}
where in the second equality we used the cycling property of the trace. This completes the proof of Eq.~\eqref{eq:friction_tensor_unsymmetric}.

\section{Positivity of the friction tensor}
\label{appendix:PSD}
In this appendix, we show that the matrix $\zeta(\lambda)$ is positive semidefinite for any $\lambda$. To do so, we will show that, up to second order in $\epsilon$, $\sum_{\mu\nu}\dot{\lambda}_{\mu}\zeta_{\mu\nu}\dot{\lambda}_{\nu} =\dot{\Sigma}_\text{na} \geq 0$. Keeping only the leading term in Eq.~\eqref{eq:excess_flux_A_expansion}, we obtain
\begin{align}
    \begin{split}
        \label{eq:psd_excess_flux}
            \dot{\Pi}_\text{ex} &= -\dv{\langle\hat{\Phi}_{\lambda}\rangle}{t} - \Tr[\dv{\log\hat{\pi}_\lambda}{t}\mathcal{L^+_\lambda}\dv{\hat{\pi}_\lambda}{t}] + \mathcal{O}(\epsilon^3)\\
            &=-\dv{\langle\hat{\Phi}_{\lambda}\rangle}{t} +\sum_{\mu\nu}\dot{\lambda}_\mu\zeta_{\mu\nu}\dot{\lambda}_\nu + \mathcal{O}(\epsilon^3)
    \end{split}
\end{align}

From Hatano-Sasa's relation [Eq.~\eqref{eq:modified_second_law}], $\dot{\Pi}_\text{ex}^{(A)} = -\dot{S} + \dot{\Sigma}_{\text{na}}$. By inserting this in Eq.~\eqref{eq:psd_excess_flux} and rearranging the result, we can see that
\begin{equation}
    \sum_{\mu\nu}\dot{\lambda}_\mu\zeta_{\mu\nu}\dot{\lambda}_\nu = \dv{t}(\langle\hat{\Phi}_{\lambda}\rangle - S(\hat{\rho}_t)) + \dot{\Sigma}_{\text{na}} + \mathcal{O}(\epsilon^3).
\end{equation}
Now, we notice that the difference $\langle\hat{\Phi}_{\lambda}\rangle - S(\hat{\rho}_t)$ is, in fact, the relative entropy, which can be seen from its definition:
\begin{equation}
    D(\hat{\rho}_t||\hat{\pi}_{\lambda}) = \Tr[\hat{\rho}_t(\log{\hat{\rho}}-\log{\hat{\pi}_\lambda})] = -S(\rho_t) + \langle\hat{\Phi}_{\lambda}\rangle.
\end{equation}
Therefore,
\begin{equation}
    \sum_{\mu\nu}\dot{\lambda}_\mu\zeta_{\mu\nu}\dot{\lambda}_\nu = \dv{t}D(\hat{\rho}||\hat{\pi}_{\lambda}) + \dot{\Sigma}_{\text{na}} + \mathcal{O}(\epsilon^3).
\end{equation}
We will now show that the derivative of the relative entropy is of order $\mathcal{O}(\epsilon^3)$, and therefore is negligible for slow processes. The relative entropy can be expanded as \cite{hiai2014introduction}
\begin{equation}
    D(\hat{\pi}_{\lambda}+\delta\hat{\rho}_t||\hat{\pi}_{\lambda})=\Tr[\delta \hat{\rho}_t\mathbb{J}_{\hat{\pi}}^{-1}(\delta \hat{\rho}_t)] + \mathcal{O}(\epsilon^3).
\end{equation}
where $\mathbb{J}_{\hat{\pi}}^{-1}$ is the super-operator
\begin{equation}
    \mathbb{J}_{\hat{\pi}}^{-1}(\cdot) = \int_0^\infty ds\, (\hat{\pi} + s\mathbb{1})^{-1} (\cdot) (\hat{\pi} + s\mathbb{1})^{-1},
\end{equation}
which is in fact the inverse of the Kubo-Mori operator,  $\mathbb{J}_{\hat{\pi}}(\cdot)=\int_0^1ds\, \hat{\pi}_\lambda^s (\cdot)\hat{\pi}_\lambda^{1-s}$. From the expansion in Eq.~\eqref{eq:expansion_rho}, we see that $\delta\hat{\rho}=\mathcal{O}(\epsilon)$, therefore $\Tr[\delta \hat{\rho}_t\mathbb{J}_{\hat{\pi}}^{-1}(\delta \hat{\rho}_t)]=\mathcal{O}(\epsilon^2)$. Taking the derivative, we have that
\begin{equation}
    \dv{t}D(\hat{\rho}||\hat{\pi}_{\lambda}) = \mathcal{O}(\epsilon^3).
\end{equation}
Therefore,
\begin{equation}
    \label{eq:action_sigma_na_appendix}
    \sum_{\mu\nu}\dot{\lambda}_\mu\zeta_{\mu\nu}\dot{\lambda}_\nu = \dot{\Sigma}_{\text{na}} + \mathcal{O}(\epsilon^3).
\end{equation}
Integrating over the path leads to
\begin{equation}
\int_0^T\mathrm{d}t\,\sum_{\mu\nu}\dot{\lambda}_\mu\zeta_{\mu\nu}\dot{\lambda}_\nu = \int_0^T\mathrm{d}t\,\dot{\Sigma}_{\text{na}} + \mathcal{O}(\epsilon^2),
\end{equation}
which is Eq.~\eqref{eq:action_metric_non_ad_ent} of the main text. Now, to prove that $\zeta(\lambda)$ is positive semidefinite, we need to show that $\sum_{\mu\nu}x_\mu\zeta_{\mu\nu}x_\nu \geq 0$ for every vector $x$. For a given $\lambda_0$, consider the linear protocol $\lambda_t = \lambda_0 + \epsilon xt$. By Eq.~\eqref{eq:action_sigma_na_appendix},
\begin{equation}
    \epsilon^2 \sum_{\mu\nu}x_\mu\zeta_{\mu\nu}(\lambda_0)x_\nu =\dot{\Sigma}_\text{na} + \mathcal{O}(\epsilon^3).
\end{equation}
Therefore, by taking $\epsilon$ small enough, we have that $\sum_{\mu\nu}x_\mu\zeta_{\mu\nu}(\lambda_0)x_\nu =\dot{\Sigma}_\text{na}\geq 0$, which completes the proof.

\section{Scaling of the adiabatic and nonadiabatic entropy production}

In this appendix, we briefly discuss the scaling behavior of the adiabatic and nonadiabatic entropy production in the slow-driving limit. 

The adiabatic contribution $\Sigma_\text{ad}$ reflects the entropy required to maintain the system in an instantaneous NESS throughout the protocol.
In the slow-driving regime, the system remains close to this steady state, and the associated entropy production rate $\dot{\Sigma}_{\text{ad}}(t)$ varies slowly in time. Integrating over the protocol duration $T$, we expect:
\begin{equation}
    \Sigma_\text{ad} \sim \mathcal{O}(T).
\end{equation}

In contrast, the nonadiabatic contribution $\Sigma_\text{na}$ arises from deviations from the steady state and vanishes in the quasistatic limit. As shown in Eq.~\eqref{eq:action_metric_non_ad_ent} of the main text, the slow-driving expansion yields
\begin{equation}
    \Sigma_\text{na} \sim \mathcal{O}\left(\frac{1}{T}\right),
\end{equation}
where we used the fact that the protocol time is inversely proportional to the driving speed, $T \sim 1/\epsilon$. 

These scalings confirm the distinct dynamical roles of the two contributions: $\Sigma_{\text{ad}}$ represents the unavoidable baseline cost of operating in a nonequilibrium steady state, while $\Sigma_{\text{na}}$ captures the additional, protocol-dependent cost of driving.

\section{Three-Level Maser}
\label{appendix:3LM}
In the following, the set-up of the three-level maser (see e.g. Ref.~\cite{Cangemi2024} for a recent review), shown in Fig.~\ref*{fig:3LM_all}a) in the main text, is discussed.
The two baths with inverse temperatures $\beta_1$ and $\beta_2$, respectively, couple the levels $\ket{g}\leftrightarrow\ket{e_A}$ and $\ket{e_B}\leftrightarrow\ket{e_A}$. The work bath is assumed to have an infinite temperature and mediates the transition $\ket{g}\leftrightarrow\ket{e_B}$. Because of the infinite temperature assumption, there is no entropy flux due to the energy flow from the central system to the respective bath. The GKLS master equation for the three-level system is given by
\begin{equation}
    \dv{\hat{\rho}}{t} = -i\left[{H}, {\hat{\rho}}\right] + \mathcal{D}_1({\hat{\rho}})+ \mathcal{D}_2({\hat{\rho}})+ \mathcal{D}_3({\hat{\rho}})
\end{equation}
with the Hamiltonian ${H} =  \epsilon_g \ketbra{g}{g} + \epsilon_{A} \ketbra{e_A}{e_A} + \epsilon_{B} \ketbra{e_B}{e_B}$ and dissipators
\begin{equation}
\begin{split}
    \mathcal{D}_x({\hat{\rho}}) =& \gamma_x \left(\hat{A}_x {\hat{\rho}}\hat{A}_x^\dagger -\frac{1}{2}\left\lbrace \hat{A}_x^\dagger \hat{A}_x, {\hat{\rho}}\right\rbrace\right)\\
    &+ \gamma_x \exp\left(-\beta_x \omega_x\right)\left(\hat{A}^\dagger_x {\hat{\rho}}\hat{A}_x -\frac{1}{2}\left\lbrace\hat{A}_x\hat{A}_x^\dagger , {\hat{\rho}}\right \rbrace\right)
\end{split}
\end{equation}
with $\hat{A}_1 = \ketbra{g}{e_A} $, $\hat{A}_2 = \ketbra{e_B}{e_A} $,  $\hat{A}_3 = \ketbra{g}{e_B} $, and $\omega_1 = \epsilon_{A} - \epsilon_g$, $\omega_2 = \epsilon_{A} - \epsilon_{B}$ and  $\omega_3 = \epsilon_{B} - \epsilon_g$.\\
The two naive protocols we consider in the main text are, given a protocol duration $T$, and the initial and target inverse temperatures $\beta_1^0$ and $\beta_1^T$, 
\begin{align}
    \beta^\text{lin}_h(t) &= \beta_1^0 + (\beta_1^T - \beta_1^0) \frac{t}{T}\\
    \beta^{\text{sin}^2}_h(t) &= \beta_1^0 + (\beta_1^T - \beta_1^0)  \sin(\frac{\hat{\pi}}{2} \frac{t}{T})^2.
\end{align}

\section{Geodesics for single-parameter protocols}
\label{appendix:single_parameter_geo}
Since we consider a single-parameter protocol, the geodesic drive $\beta_1(t)$, given a finite protocol duration $T$, can be obtained as follows:
We know that the speed along the geodesic $v_\text{geo}$ is constant. This means that the total statistical distance traveled along the geodesic path is $\ell^\mathrm{geo}_T = v_\text{geo} T$, so that
\begin{equation}
    \ell^\mathrm{geo}_T = \int_0^T dt \dv{\beta_1}{t} \sqrt{m_{\beta_1}} = \int_{\beta_1^0}^{\beta_1^T} d\beta_1\sqrt{m_{\beta_1}} = v_\text{geo} T.
\end{equation}
Therefore, we can obtain $v_\text{geo} = \ell^\mathrm{geo}_T/T$. This yields the inverse function of what we would like to obtain: $t(\beta_1)$, for any $0\leq t\leq T$
\begin{equation}
    t(\beta_1) = \frac{1}{v_\text{geo}}\int_{\beta_1^0}^{\beta_1} d\tilde\beta_1\sqrt{m_{\tilde\beta_1}}. 
\end{equation}
By inverting this function, we obtain the geodesic drive $\beta_1(t)$, without having to solve the geodesic equation.\\
For the example of the three-level maser, we find the analytical form of the metric as a function of the control parameter $\beta_1$
 \begin{align}
 m_{\beta_1} &= \frac{\omega_1^2 e^{2 \beta_1(\epsilon_g+\epsilon_{A})} \left(e^{\beta_2\epsilon_{B}}+e^{\beta_2\epsilon_{A}}\right)}{\gamma  \left(e^{\beta_1\epsilon_g}+2 e^{\beta_1\epsilon_{A}}\right)^3 \left(e^{\beta_1\epsilon_g+\beta_2\epsilon_{B}}+e^{\beta_1\epsilon_g+\beta_2\epsilon_{A}}+e^{\beta_2\epsilon_{B}+\beta_1\epsilon_{A}}\right)},\\
     I_{\beta_1}^\mathrm{KMB}(\hat{\pi}_{\beta_1})&= \frac{\omega_1^2 e^{ (2 \epsilon_g+\epsilon_{A})} \left(e^{\beta_2 \epsilon_{B}}+e^{\beta_2 \epsilon_{A}}\right)}{\left(e^{\beta_1 \epsilon_g}+2 e^{\beta_1 \epsilon_{A}}\right)^2 \left(e^{\beta_1 \epsilon_g+\beta_2 \epsilon_{B}}+e^{\beta_1 \epsilon_g+\beta_2 \epsilon_{A}}+e^{\beta_2 \epsilon_{B}+\beta_1 \epsilon_{A}}\right)},\\
     \tau_{\beta_1}&= \frac{1}{ \gamma (2+  e^{-\beta_1 \omega_1})},
 \end{align}
which factorizes into the generalized relaxation time $\tau_{\beta_1}$ and the KMB quantum Fisher metric of the the instantaneous steady state $ I_{\beta_1}^\mathrm{KMB}$, as discussed in the main text.
Interestingly, the integral relaxation time depends solely on $\gamma$, which sets the time scale of the dynamics, and the Boltzmann factor $e^{-\beta_1 \omega_1}$ associated with transitions in the central system induced by interactions with the bath.
In the infinite temperature limit, $\beta_1\to 0$, $\tau_{0} = \frac{1}{3\gamma}$, whereas in the zero temperature limit, $\beta_1\to \infty$,  $\tau_{\infty} = \frac{1}{2\gamma}$.
\section{Quantum Fisher information with respect to time}
\label{appendix:FI_time}
When discussing (quantum) Fisher information, one typically considers a finite set of parameters $\theta = (\theta_1, \dots, \theta_M)$ that are encoded into the state of a system. The manifold of states with parameters as coordinates is equipped with a metric tensor $m_{ij}$, defining the distance measure between states
\begin{equation}
ds^2 = \frac{1}{4} \sum_{ij} m_{ij} \, d\theta_i d\theta_j.
\end{equation}
The metric $m_{ij}$ is known as the (quantum) Fisher information matrix.
One may alternatively treat time itself as a  parameter, and then the (quantum) Fisher information becomes
\begin{equation}
F(t) = \sum_{ij} \frac{d\theta_i}{dt} m_{ij} \frac{d\theta_j}{dt}
\end{equation}
again providing a metric $ds^2 = \frac{1}{4} F(t) dt^2$ that can be employed to define a  statistical distance.

Let us first consider a classical scenario of a discrete state system with a finite number $M$ of configurations. The system state can then be described by the discrete set of occupation probabilities $\{p_x(t)\}$ with $x = 1, \dots, M$ ($p_x(t) \geq 0$ and $\sum_x p_x(t) = 1$), and in that case the metric is uniquely defined by  $m = \sum_x p_x(t) \left(\frac{\partial \log p_x(t)}{\partial t}\right)^2$~\cite{cencov_statistical_2000}. The significance of the classical Fisher information with respect to time for stochastic thermodynamics is discussed in Refs.~\cite{NicholsonPRE2018, ItoPRL2018}.
In quantum dynamics, the quantum Fisher information (QFI) with respect to time has also gained
significant interest, particularly in developing quantum speed limits~\cite{marvian_operational_2022, girolami_how_2019, GarciaPRX2022} and due to its connection to stochastic thermodynamics~\cite{BettmannPRE2024}. The QFI in quantum systems, defined on a manifold of density matrices, generalizes the classical Fisher information. It is characterized by a family of metrics, governed by the Morozova, \v{C}encov, and Petz theorems~\cite{petz_introduction_2011, petz_riemannian_1996}. Any metric contractive under stochastic maps, serving as faithful measures of distinguishability in quantum state space~\cite{scandi_quantum_2024}, must yield a squared line element of the form
\begin{equation}
ds^2 = \frac{1}{4} \sum_{x, y} \frac{|d\hat{\rho}_{xy}|^2}{p_x f(p_y/p_x)},
\end{equation}
where $\hat{\rho} = \sum_x p_x \ket{x}\bra{x}$ is the density matrix, $d\hat{\rho}_{xy} := \bra{x}d\hat{\rho}\ket{y}$, and $f$ is an operator monotone, self-inversive function satisfying $f(1) = 1$.
The infinitesimal line element is given by
\begin{equation}
ds^2 = \frac{1}{4} \sum_{x, y} \frac{|\partial_t \hat{\rho}_{xy}(t)|^2}{p_x(t) f(p_y(t)/p_x(t))} \, dt^2,
\end{equation}
and th quantum Fisher information $F_Q(t)$ is
\begin{equation}
F_Q(t) = \sum_{x, y} \frac{|\partial_t \hat{\rho}_{xy}(t)|^2}{p_x(t) f(p_y(t)/p_x(t))}.
\end{equation}
Prominent examples of QFIs include the symmetric logarithmic derivative (SLD) QFI, the Wigner-Yanase (WY) QFI, Kubo-Mori-Bogoliubov (KMB) QFI and the harmonic mean (HM) QFI, with different functions $f$, see Ref.~\cite{scandi_quantum_2024} for a recent review.


\end{document}